\documentclass[prl,showpacs,floatfix,twocolumn,superscriptaddress]{revtex4}
\usepackage{graphicx}
\usepackage{dcolumn}

\begin{document}

\newcommand*{\cm}{cm$^{-1}$\,}
\newcommand*{\tise}{1$T$-TiSe$_2$\,}

%
\title{Semimetal to semimetal charge density wave transition in 1\textit{T}-TiSe$_2$}
%
%
\author{G. Li}
\author{W. Z. Hu}
\affiliation{Beijing National Laboratory for Condensed Matter
Physics, Institute of Physics, Chinese Academy of Sciences,
Beijing 100080, P. R. China}
\author{D. Qian}
\author{D. Hsieh}
\author{M. Z. Hasan}
\affiliation{Department of Physics, Joseph Henry Laboratories of
Physics, Princeton University, Princeton, New Jersey 08544, USA}

\author{E. Morosan}
\author{R. J. Cava}
\affiliation{Department of Chemistry, Princeton University,
Princeton, New Jersey 08540, USA}


%
\author{N. L. Wang}
\email{nlwang@aphy.iphy.ac.cn}%
\affiliation{Beijing National Laboratory for Condensed Matter
Physics, Institute of Physics, Chinese Academy of Sciences,
Beijing 100080, P. R. China}
%
%
%

\begin{abstract}
We report an infrared study on 1$T$-TiSe$_2$, the parent compound
of the newly discovered superconductor Cu$_x$TiSe$_2$. Previous
studies of this compound have not conclusively resolved whether it
is a semimetal or a semiconductor: information that is important
in determining the origin of its unconventional CDW transition.
Here we present optical spectroscopy results that clearly reveal
that the compound is metallic in both the high-temperature normal
phase and the low-temperature CDW phase. The carrier scattering
rate is dramatically different in the normal and CDW phases and
the carrier density is found to change with temperature. We
conclude that the observed properties can be explained within the
scenario of an Overhauser-type CDW mechanism.
\end{abstract}


\maketitle

%

Charge density waves (CDW) and superconductivity are two important
collective phenomena in solids. There has been tremendous interest
in the interplay between these two states in condensed matter
physics. The recent discovery of superconductivity upon controlled
intercalation of Cu into CDW-bearing \tise \cite{Morosan} offers a
good opportunity to study the transition between the two states.
As a first step, understanding the parent compound \tise is
essential.

The CDW state in \tise has been known for several decades, but
remains poorly understood. It shows a lattice instability around
200 K, below which it enters into a commensurate CDW phase
associated with a ($2\times2\times2$)
superlattice\cite{Wilson1,Wilson2,Salvo}. The electronic structure
has been studied by band structure calculations\cite{Zunger} and
angular resolved photoemission spectroscopy (ARPES)
measurements\cite{Bachrach,Traum,Anderson,Pillo,Kidd,Rossnagel,Cui}.
At energies near E$_F$, \tise is governed by two bands: a
hole-like Se 4p band close to the $\Gamma$ point and a Ti 3d band
around the $L$ point. Due to the small energy differences
involved, ARPES experiments have not unambiguously shown
 whether the compound is a \textit{semiconductor} or a
\textit{semimetal} in either the normal or CDW states.

This issue is central to resolving the mechanism of the CDW phase
transition, which remains problematic. Contrary to other 1$T$-type
transition metal dichalcogenides, the CDW transition in
1$T$-TiSe$_2$ is not likely to originate from Fermi surface (FS)
nesting or saddle-point singularities, as parallel FS sheets have
neither been predicted\cite{Zunger} nor
observed\cite{Bachrach,Traum,Anderson,Pillo,Kidd,Rossnagel,Cui,Qian}.
One picture for the driving force for CDW formation is exciton
formation at low temperature. However, the available experimental
results in support of this picture are
controversial\cite{Wilson3,Traum,Anderson,Pillo,Kidd,Rossnagel}.
There exist several other proposals for the driving mechanism,
including an indirect Jahn-Teller effect\cite{Kidd,Snow}, but no
conclusive agreement has been reached. In this Letter, we study
the change of the electronic structure as a function of
temperature by optical spectroscopy. Although there exist several
optical studies on this compound, most of them were performed only
at room temperature or in the high energy region (above $\sim~$0.5
eV)\cite{Wilson1,Beal,Holy,Hugues,Wilson4,Bayliss}. A detailed
study to low energy has been lacking. We argue here that
Overhauser's scenario for CDW formation due to electron-hole bound
pairs offers the most likely explanation to account for all the
observed features.

Single crystals with reflective flat surfaces were grown as
previously described \cite{Morosan2}. The dc resistivity was
measured by a standard four probe technique in a Quantum Design
PPMS. Similar to the data reported earlier\cite{Cui}, the
$\rho(T)$ curve, as displayed in the inset of Fig. 1(a), shows a
rather striking shape. The resistivity increases with decreasing T
from 300 K and shows no distinct anomaly at the CDW transition
near 200 K. After reaching a maximum near 150 K, it decreases on
further cooling. This behavior is rarely seen in metallic or
semiconducting compounds, and is also not seen in conventional
CDW-bearing materials.

The frequency-dependent reflectance spectra R($\omega$) at
different T were measured by a Bruker IFS 66 v/s spectrometer in
the range from 50 to 25,000 cm$^{-1}$ and a grating-type
spectrometer from 20,000 to 50,000 \cm. The sample was mounted on
an optically black cone on the cold-finger of a He flow cryostat.
An \emph{in situ} gold (50 $\sim$ 15,000 cm$^{-1}$) and aluminum
(9,000 $\sim$ 50,000 cm$^{-1}$) overcoating technique was employed
for reflectance measurements. A Kramers-Kronig transformation of
R($\omega$) was used to obtain the other optical response
functions. A Hagen-Rubens relation was used for low frequency
extrapolation, and a constant extrapolation for high-frequency to
300,000 cm$^{-1}$ followed by an $\omega^{-2}$ function was used
for the higher energy side.

\begin{figure}[t]
\centerline{\includegraphics[width=2.8in]{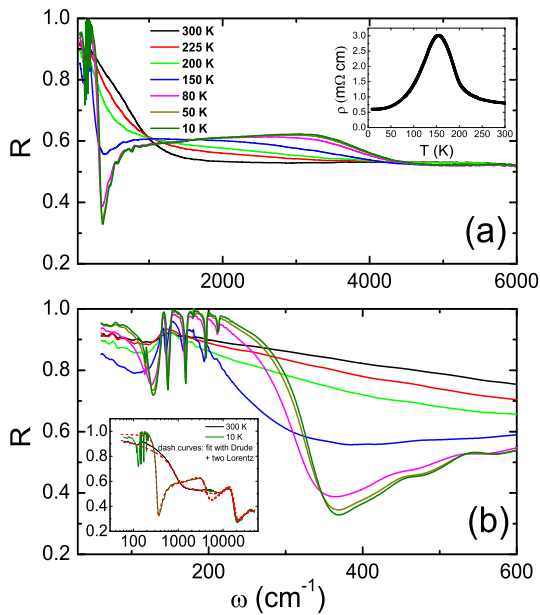}}%
\vspace*{-0.20cm}%
\caption{(Color online) (a) The temperature dependent R($\omega$)
in the frequency range from 30 \cm to 6,000 \cm. The inset is the
T-dependent dc resistivity. (b) The expanded plot of R($\omega$)
from 30 \cm to 600 \cm. The inset shows
R($\omega$) data up to 50,000 \cm on a logarithm scale. Dash curves are fitted to a simple Drude-Lorentz model.}%
\label{fig1}
\end{figure}

Fig. 1 (a) shows the R($\omega$) spectra at different T over the
frequency range from 50 \cm to 6,000 \cm and Fig. 1 (b) shows the
spectra in an expanded plot from 50 to 600 \cm. The inset shows
the spectra over from 50 to 50,000 \cm on a logarithmic scale at
two different temperatures. There is a very strong phonon
structure near 130 \cm at both high and low temperatures. The
antiresonance behavior of this phonon indicates strong coupling
between electrons and the phonon. In the CDW phase below 200 K,
several new phonon lines appear, evidencing the symmetry change of
the structure.

Most importantly, the reflectance spectra show an unambiguously
metallic character at both high and low T. The carrier damping
rate is dramatically different in the normal and CDW phases. The
R($\omega$) at high T displays a linear-$\omega$ dependence up to
an edge of 1,100 \cm. This is the well-known overdamped behavior,
similar to the situation in high-T$_c$ cuprates, indicating rather
strong carrier scattering. With decreasing T, R($\omega$) at low
$\omega$ decreases, leading to an edge shift towards the low
energy region. At the same time, R($\omega$) between 1,000 and
5,000 \cm increases. When the sample is cooled below 150 K, a
sharp, well-defined plasma edge appears and then shifts slightly
to higher $\omega$ with decreasing T. The screened plasma edge
frequency is located between 300-350 \cm ($\sim$40 meV) at low T,
which is almost two orders of magnitude lower in frequency than
for usual metals. The extremely low plasma edge frequency reveals
a very small carrier density. Such a well-defined plasma edge is
an indication of rather small carrier damping.

\begin{figure}[t]
\centerline{\includegraphics[width=2.8in]{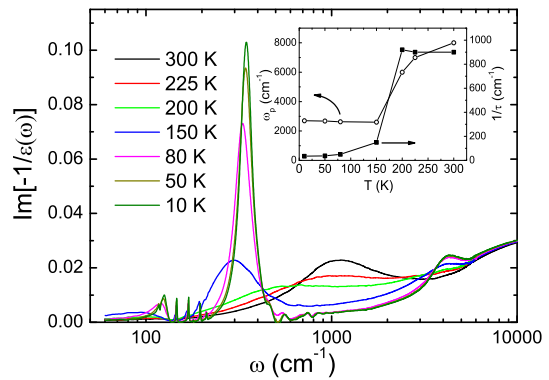}}%
\vspace*{-0.20cm}%
\caption{(Color online) The energy loss function spectra at
different T. Inset: the variation of the plasma frequency and
scattering rate for the Drude term in Equ. (1).}%
\label{fig2}
\end{figure}

The formation of the plasma edge and its evolution with T can be
seen more clearly from the energy loss function,
$Im{[-1/\epsilon(\omega)]}$, shown in Fig. 2. In the energy loss
function, the peak position indicates the screened plasma
frequency, while the peak width is related to the carrier damping
rate. As T decreases from 300 K to 150 K, the screened plasma
frequency decreases from 1132 \cm to 300 \cm, but it then
increases from 300 \cm to 347 \cm as the compound is cooled from
150 K to 10 K. At the same time, the spectral shape changes from a
broad feature to a sharp peak structure, suggesting a dramatic
reduction of the carrier damping at low T. The peak-like feature
near 4,000 \cm is not the plasma edge, but is the interband
transition due to the formation of a gap.

To quantitatively analyze the carrier density and damping, we fit
the experimental reflectance to a simple Drude-Lorentz model:
\begin{equation}
\epsilon(\omega)=\epsilon_\infty-{{\omega_p^2}\over{\omega^2+i\omega/\tau}}+\sum_{i=1}^2{{S_i^2}\over{\omega_i^2-\omega^2-i\omega/\tau_i}}.
\label{chik}
\end{equation}
This expression contains a Drude term and two Lorentz terms, which
approximately capture the contributions by free carriers and
interband transitions. As shown in the inset of Fig. 1(b), this
simple model can reasonably reproduce the $R(\omega)$ curves: the
calculated curves at 300 K and 10 K are displayed. In the inset of
Fig. 2, we plot the plasma frequency $\omega_p$ and scattering
rate $1/\tau$ obtained for the Drude term as a function of T. Both
parameters show a dramatic change in the temperature range of
150-200 K, in conjunction with the CDW phase transition. An
$\epsilon_\infty$=19 is obtained by fitting. (As the interband
transitions are not our focus here, we do not show the parameters
for the Lorentz terms.) From this approach, we obtain the
unscreened plasma frequency $\omega_p$$\sim$8000 \cm at 300 K,
which gives a carrier density of 7.1$\times$10$^{20}$ 1/cm$^3$ if
m=m$_e$ is assumed, which is in good agreement with band structure
calculation\cite{Zunger}. At 10 K, the value of
$\omega_p$$\sim$2700 cm$^{-1}$ implies a carrier density of
8.1$\times$10$^{19}$ 1/cm$^3$.

\begin{figure}[t]
\centerline{\includegraphics[width=2.7in]{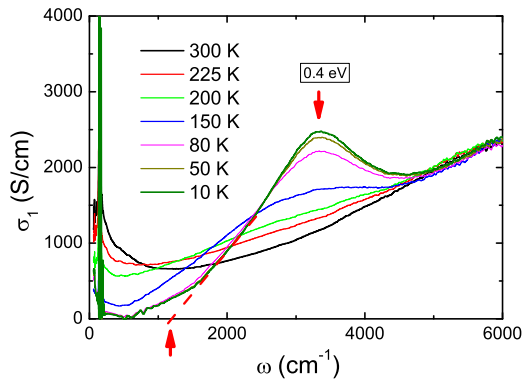}}%
\vspace*{-0.20cm}%
\caption{(Color online) Frequency dependence of the optical conductivity at different temperatures.}%
\label{fig3}
\end{figure}

Figure 3 shows the optical conductivity $\sigma_1(\omega$) at
different T below 6,000 \cm. There are two important features. One
is the dramatic suppression of the spectral weight at low $\omega$
below 200 K. The lost spectral weight roughly shifts to the region
between 2,000 \cm and 5,000 \cm, leading to a very pronounced peak
centered at 3,200 \cm (0.4 eV). This is a clear indication of the
formation of a gap in the charge excitation spectrum. The gap
value can be roughly defined by the linear extrapolation of the
peak edge, about 1,250 \cm (0.15 eV) at lowest T. The second is
the Drude-like response at both high and low T, if the very strong
phonon contribution is subtracted. The conductivity value at the
low frequency limit, which decreases as T decreases from 300 K to
150 K but increases from 150 K to 10 K, matches well the
T-dependent behavior of the dc resistivity. A very narrow, sharp
Drude component develops below the gap frequency in the CDW phase.

\begin{figure}[t]
\centerline{\includegraphics[width=2.3in]{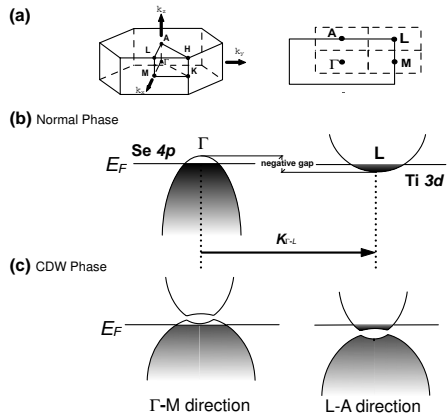}}%
\vspace*{-0.20cm}%
\caption{ Schematic picture for the evolution of the band
structure with temperature. (a) The Brillouin zone layout for
\tise. The right part shows the new Brillouin zones (dashed lines)
for a $2\times2\times2$ superlattice in $\Gamma, A, L$ and $M$
plane. (b) The semimetal picture in the normal phase. A small
number of Se $4p$ holes and Ti $3d$ electrons exist near $\Gamma$
and $L$, respectively. (c) The exciton state in the CDW phase
caused by the Coulomb interaction between electrons and holes.}%
\label{fig4}
\end{figure}

Previous experiments have not conclusively shown whether \tise is
a semiconductor with a very small energy gap or a semimetal with
very low carrier density. As a bulk probe with very fine energy
resolution, the present optical measurements reveal clearly that
the compound is a semimetal with very low carrier density both
above and below the CDW transition. Our data, in combination with
results obtained from ARPES experiments, lead us to propose a
schematic picture for the evolution of the electronic structure
with T, displayed in Fig. 4. At room T, small numbers of Se 4$p$
holes $n_h$ and Ti 3$d$ electrons $n_e$ exist near $\Gamma$ and
$L$, respectively. Charge neutrality based on the stoichiometry of
\tise requires the same number of holes and electrons. This
semimetallic picture, shown in Fig. 4(b), is in agreement with the
band structure calculation by Zunger\cite{Zunger}. In this case,
the plasma frequency reflects the contribution of two kinds of
carriers with $\omega_p^2=4\pi e^2({n_h\over m_h}+{n_e\over
m_e})$, where $m_h$ and $m_e$ are the effective mass of $p$ hole
and $d$ electrons, respectively.

On decreasing T from room temperature, the carrier density is
expected to decrease due to the reduction of the thermally excited
carriers for such a semimetal with small negative gap. This is the
reason why the plasma edge shifts to lower $\omega$. The increase
of the dc resistivity in this temperature regime can be ascribed
to this reduction of carrier numbers. An important consequence of
the carrier number reduction is that the Coulomb interaction
between electrons and holes becomes poorly screened. When the
carrier density is low enough, the Coulomb interaction can lead to
a bound state between an electron and a hole,\textit{ i.e.} an
exciton. As a result, the free carriers are largely removed from
the FS below 200K.

Kohn and others have proposed an exciton driving mechanism for a
CDW instability in a semiconductor or semimetal with a small
indirect energy gap\cite{Kohn,Halperin}. In the semimetal case,
the electron-hole coupling acts to mix electron bands and hole
bands that are connected by particular wave vectors which match
the superlattice of CDW phase. The associated CDW transition,
driven by the energy gain that arises from the gapping of the FS,
is referred to as Overhauser type\cite{Kohn,Overhauser}.
Considering charge neutrality, it appears that a direct mixing of
the hole band near $\Gamma$ and electron band near $L$ can only
lead to a hybridization gap at E$_F$, and thus an insulator in the
CDW phase, in contradiction to experiments. \tise, however,
appears to be a very special case. Because the ordering wave
vector is commensurate, $2\times2\times2$, the new Brillouin zone
is reduced in all three directions. Then, all the four points,
$\Gamma, A, L$ and $M$ shown in Fig. 4 (a), become new zone center
of the $2\times2\times2$ superlattice. In this case, we need to
consider all possible band mixing associated with the superlattice
of CDW phase. A band anisotropy can clearly be identified in the
band structure calculations\cite{Zunger}, which is an essential
ingredient for such a semimetal-semimetal transition \cite{Kohn}.
In addition to the hole and electron bands that cross E$_F$ near
$\Gamma$ and $L$, respectively, the Ti 3d band near $M$ is
slightly higher than E$_F$, and the Se 4p band near $A$ is
slightly lower than E$_F$\cite{Zunger}. Apparently, the mixing of
the band near $M$ with that near $\Gamma$ leads to a hole-like
sub-band (the split upper part of bands may not touch E$_F$),
while the mixing of the band near $A$ with that near $L$ leads to
an electron-like band (Fig. 4(c)). Therefore, the compound enters
into another semimetal state with a new balance of holes and
electrons below the CDW transition. As indicated by Kohn, the new
hole and electron bands have to be along different k directions.
This is precisely what occurs for \tise. The above picture can
explain very naturally the coexistence of the gap opening (due to
the $\Gamma$-$L$ band mixing) and metallic conduction in the CDW
phase.

In the CDW state (below 200 K) the scattering of carriers is
strongly reduced, leading to the rapid sharpening of the plasma
edge in R($\omega$). The reduction of scattering can be attributed
to the loss of a scattering channel by FS gapping. The sharp
reduction of the conducting carrier density and the sharp
reduction of carrier scattering leads to a peak in the dc
resistivity near 150 K. Below 150 K, the carrier numbers do not
further decrease, instead they increase slightly. The reduction of
the dc resistivity on cooling below 150K is dominated by the small
and further decreasing carrier scattering rate.

The Overhauser-type CDW picture driven by strong electron-hole
coupling appears to provide a good explanation for the variation
of the dc and optical transport properties through the CDW
transition in \tise. Excitons in semiconductors are a commonly
observed phenomenon. However, such excitons are usually created by
shining light on the semiconductor, which creates an equal number
of electrons and holes. Such optically generated excitons have a
very short life time, decaying quickly via the emission of light.
The situation in \tise is very different. It has an equal number
of electrons and holes in \textit{\textbf{K}} space in different
bands. The very small carrier density leads to a stable formation
of excitons at low T. \tise might be the only real material for
which the CDW transition is realized by the exciton condensation
mechanism proposed by Kohn and
others\cite{Kohn,Halperin,Overhauser}. A closely analogous
phenomenon is the Bose condensation of an electron-hole bound
state (exciton) realized in connection with the quantum Hall
effect observed in semiconducting bilayer
systems\cite{Eisenstein}.

To summarize, the optical measurements reveal that: (1) \tise is a
semimetal with very low carrier density both above and below the
CDW transition temperature; (2) the carrier density changes with
temperature, decreasing from room T to roughly 150 K, and then
increases slightly with further decreasing T; (3) an energy gap
$\sim$0.15 eV develops below 200 K in the CDW phase; and (4)
dramatically different carrier scattering rates are present in the
normal and CDW phases. The experimental data strongly suggest the
presence of an Overhauser-type CDW transition in \tise. Further
study will help to understand why such an unusual CDW-bearing
parent compound becomes a superconductor when a small percentage
of excess electrons are doped into it via Cu intercalation.

We acknowledge very helpful discussions with W. Ku, T. M. Rice, T.
Tohyama, R.S. Markiewicz, Z. Q. Wang, X. C. Xie, and Lu Yu. This
work is supported by the National Science Foundation of China, the
Knowledge Innovation Project of Chinese Academy of Sciences, and
the Ministry of Science and Technology of China (973 project No.
2006CB601002). MZH and RJC acknowledge partial support through the
NSF(DMR-0213706) and U.S.DOE/DE-FG-02-05ER46200 and
DE-FG02-98ER45706.

%
%

\end{document}